# Surface and Interface Properties of $La_{2/3}Sr_{1/3}MnO_3$ Thin Films on $SrTiO_3$ (001)

Lina Chen[1&], Zhen Wang[1,2], Gaomin Wang[1], Hangwen Guo[1#], Mohammad Saghayezhian[1], Zhaoliang Liao[1,*], Yimei Zhu[2], E.W. Plummer[1], and Jiandi Zhang[1]

[1]*Department of Physics and Astronomy, Louisiana State University, Baton Rouge, LA 70803, USA.*

[2]*Condensed Matter Physics and Materials Science Department, Brookhaven National Laboratory, Upton, NY 11973, USA*

## ABSTRACT

Understanding and manipulating properties emerging at a surface or an interface require a thorough knowledge of structure-property relationships. We report a study of a prototype oxide system, $La_{2/3}Sr_{1/3}MnO_3$ grown on $SrTiO_3$(001), by combining *in-situ* angle-resolved x-ray photoelectron spectroscopy, *ex-situ* x-ray diffraction, and scanning transmission electron microscopy/spectroscopy with electric transport measurements. We find that $La_{2/3}Sr_{1/3}MnO_3$ films thicker than 20 unit cells (u.c.) exhibit a universal behavior with no more than one u.c. intermixing at the interface but at least 3 u.c. of Sr segregation near the surface which is (La/Sr)O terminated. The conductivity vs film thickness shows the existence of nonmetallic layers with thickness ~ $6.5 \pm 0.9$ u.c., which is independent of film thickness but mainly relates to the deviation of Sr concentration near the surface region. Below 20 u.c., the surface of the films appears mixed (La/Sr)O with $MnO_2$ termination. Decreasing film thickness to less than 10 u.c. leads to the enhanced deviation of chemical composition in the films and eventually drives the film insulating. Our observation offers a natural explanation for the thickness-driven metal-nonmetal transition in thin films based on the variation of film stoichiometry.

PACS numbers: 75.47.-m, 68.37.-d, 68.35.Dv, 68.35.Fx,

[&]Current address: School of Physics, Nanjing University, Nanjing, P. R. China
[#]Current address: School of Physics, Fudan University, Shanghai, P. R. China
[*]Current address: University of Science and Technology of China Hefei, Anhui 230029, P. R. China

**I. Introduction**

It is common that a surface or an interface can have fundamentally different physical properties from the corresponding bulk. Examples span a wide range of phenomena: metallic surface states on simple semiconductors like Si [1] and complex oxide $SrTiO_3$ (STO) [2], surface phases of ruthenates [3-5], 2D electron gas at an interface [6], or coexisting superconducting/ferromagnetic [7-9] interface states in $LaAlO_3/SrTiO_3$. Understanding the origin of surface or interface properties necessitates a thorough knowledge of spatially resolved chemical composition and structure-property relationships. This is more essential in artificially structured multi-component oxide materials. Surface termination, lattice structure, and chemical composition, as well as imperfections like defects, are crucially important for emergent properties. Understanding the properties at an interface is challenging, as can be seen in the ongoing debates around $LaAlO_3/SrTiO_3$, where many proposed mechanisms for the observed interface phenomena are still hotly contested because of interface structure/composition issues, such as polarity effect [10], cationic mixing [11, 12], defects [13, 14], and thickness-dependent polar distortion [15].

A long-standing issue is the nature of insulating phenomenon in metallic oxide materials in ultrathin films, i.e. "*dead*" layer behavior. Many thin films of metallic oxides, such as undoped $SrVO_3$ [16], $SrRuO_3$ [17], and $LaNiO_3$ [18,19], as well as many doped $La_{1-x}Sr_xMnO_3$ [20,21], exhibit degraded metallicity with decreasing film thickness and eventually become insulating below a certain critical thickness. For $La_{1-x}Sr_xMnO_3$, several mechanisms for such as a dimensionality-driven metal-insulator transition (MIT) have been proposed: enhanced electron-electron correlations [9,17,22], change of electronic configuration [20,23,24], or interface induced effects (competing bond re-hybridization and structural mismatch) [25]. However, a clear understanding of what drives the insulating *"dead"* layer behavior is still elusive.

One outstanding issue associated with the apparent "dead' layer behavior is the variation of chemical composition. In $La_{1-x}Sr_xMnO_3$ films, Sr segregation at surface and interface has been suggested based on the spectroscopy with different tools such as X-ray photoelectron spectroscopy (XPS) and scanning transmission electron microscopy (STEM) [26-34]. Segregation was first discovered in metal alloys several centuries ago [35]. The grain boundary segregation has been extensively studied because the properties of the alloys can be dramatically affected by the segregation [36,37]. Similarly, segregation is also present at the surface and interface of films, leading to significant changes in chemical composition between the surface/interface as well as the inside of doped oxide films. This has been addressed recently by both experimental and theoretical studies [34, 38-44]. Consequently, the deviation of chemical composition dramatically alters the physics properties of the films, such as electric conductivity [45-47], structural and optical properties [48,49], polarization and magnetic ordering [50-52]. Compared with thick films, the deviation from chemical stoichiometry of the ultrathin films is much more severe, which has been suggested as origin of the dead layer formation of the ultrathin films [48, 53-57]. However, there is almost no *in-situ* characterization on the surface segregation, thus raising an issue about the effect of surface contamination, etc.

Here we demonstrate that, in $La_{2/3}Sr_{1/3}MnO_3$ (LSMO) thin films, both Sr segregation at the surface and intermixing at the interface occur in the film growth process, resulting in persistent nonmetallic layers in the films. Although Sr segregation at surface and interface have been previously suggested [26-34], we provide a quantitative study on the layer-by-layer chemical composition of LSMO films, which is critical to understand the nature of dead layer formation as well as the growth mechanism of these complex oxide films. For LSMO films that are relatively thick ($\geq$ 20 u.c.), the surface and interfacial properties are independent of film thickness. Decreasing film thickness to less than 10 u.c. further enhances the off-stoichiometry and eventually drives the films insulating.

**II. Experimental Details**

The LSMO films were grown on atomically flat $TiO_2$ terminated STO (001) substrates (non-doped and 0.1 wt% Nb-doped) using ultrahigh vacuum pulsed laser deposition [58]. Doped substrates were used to avoid charging issues for *in-situ* surface characterization such as low-energy electron diffraction (LEED) and angle-resolved X-ray photoelectron spectroscopy (ARXPS). Non-doped substrates used for *ex-situ* electrical resistivity and magnetic properties measurements. Different growth conditions, including changing oxidant background gas, growth temperature, laser operation condition, and so on, have been undertaken [58]. The growth condition for the films studied here has been optimized to obtain high quality films with minimized "dead" layer thickness. A KrF excimer laser ($\lambda$ = 248 nm) with a repetition rate of 3 Hz and a laser flounce of ~ 1 J/cm$^2$ was used for the growth. To obtain a stoichiometric film, an oxidant background gas (99% $O_2$ + 1% $O_3$) with a pressure of 80 mTorr was utilized. During growth, the substrate was maintained at 700 °C. Film growth and its thickness were monitored using *in-situ* reflection high-energy electron diffraction (RHEED). The film thickness is obtained by counting the number of oscillations with one complete oscillation corresponding to one u.c. thickness. LEED measurements were performed *in-situ* to study the surface structure. X-ray diffraction (XRD) was used to measure thin film structure.

The atomically resolved lattice structure, composition, and valence state, were determined by cross-sectional scanning transmission electron microscopy (STEM) and spectroscopy. STEM samples were cut into 50 nm thick pieces by a focused ion beam with $Ga^+$ ion milling, and then nano-milled with $Ar^+$ ions to reduce surface damage and to further thin the samples to about 30 nm. All the samples were studied using a double-aberration-corrected 200 kV JEOL ARM microscope. Two types of imaging modes were utilized: high-angle annular dark-field (HAADF) imaging, in which the intensity is proportional to the atomic number of heavy elements [59], and annular bright-field (ABF) imaging, which is sensitive to light elements such as oxygen. To quantitatively analyze lattice parameters and octahedral distortions in LSMO/STO, we measured the positions of atomic columns from the HAADF- and intensity reversed ABF- images by fitting with 2D Gaussian peak profiles.

To perform electron energy loss spectroscopy (EELS) with STEM, the imaging conditions were optimized with a probe size of 0.8 Å, a convergence semi-angle of 20 mrad, and a collection semi-angle of 88 mrad. EELS spectra mapping was obtained across the interface with a step size of 0.12 Å and a dwell time of 0.05 s/pixel. After background subtraction with a power-law function and correction by a Fourier deconvolution method for removing the multi-scattering effects, EELS intensity profiles were used to determine the elemental concentration with a calibration from a standard sample ($La_{2/3}Sr_{1/3}MnO_3$ single crystal).

ARXPS was utilized to determine the chemical composition near the surface. The core level spectra of Mn 2p, Sr 3d, and La 4d were *in-situ* measured by using a monochromated Al $K_\alpha$ X-ray source and PHOIBOS 150 energy analyzer, both from SPECS. The energy analyzer was calibrated with the core level of single crystalline gold (Au $4f_{7/2}$ peak). The depth profile of chemical components can be extracted from the ARXPS data knowing the relative photoionization cross-sections and mean free paths [60].

Resistivity as a function of temperature [$\rho(T)$] was measured *ex-situ* using a physical properties measurement system (PPMS) with a standard four-probe method. The magnetic properties of the films were determined with a Quantum Design Superconducting Quantum Interference Device (SQUID), and the results are the same as those reported previously [58]. Except for RHEED diffraction, *T*-dependent transport and magnetic properties, all data were taken when the samples were at room temperature.

## III. Results and Discussion

Figure 1(a) illustrates the ideal cubic-like structural model of LSMO films grown on a $TiO_2$-terminated STO (001) substrate. The thin films were grown in a layer-by-layer mode, which was monitored by RHEED diffraction spot oscillations, as shown in Fig. 1(b) and (c). LEED images [see Fig. 1(d)] show no fractional spots, indicating that all films maintain well-ordered $p(1\times1)$ surface structure.

The film structure was first characterized by thin film XRD showing that all films have orthorhombic symmetry. Figure 1(e) presents a coupled ($\theta$ - $2\theta$) XRD spectrum around substrate [002] diffraction peak for a 40 u.c. LSMO film. The presence of Fresnel oscillations confirms the smoothness of the surface and interface as well as the high quality of the thin film on a macroscopic level. The STO substrate has a simple cubic "$a^0a^0a^0$" structure without $TiO_6$ octahedral tilt or rotation. Bulk LSMO has a pseudo cubic (rhombohedral "$a^-a^-a^-$") structure with $MnO_6$ octahedral rotation and tilt. However, our XRD data with reciprocal lattice mappings (RLMs) shows that the film has orthorhombic ("$a^-a^-c^0$") structure. Figure 1(f) shows one RLM around the substrate cubic [-103] spot. The vertical ($Q_\parallel$) alignment of substrate and film reflections indicate that the film is fully strained with respect to the substrate. This will strongly suppress the rotational distortion of $MnO_6$ about (001) direction. The film has an averaged out-of-plane (OOP) lattice constant (also the u.c. thickness) of $c = 3.848 \pm 0.002$ Å as schematically shown in Fig. 1(g). This value is slightly smaller than the bulk pseudocubic lattice constant (3.876 Å) due to tensile strain originating from the lattice mismatch with STO (3.904 Å). It was also found that the film has an OOP tilt distortion characterized by the tilt angle ($\alpha$) [see Fig. 1(g)], which is the rotation of the octahedron about the [1-10] direction, will be discussed in detail later.

Figure 1(h) displays the *T*-dependence of resistivity for different film thickness. For thick films (> 7 u.c.), $\rho(T)$ exhibits a $T$-driven MIT coupled with a ferromagnetic to paramagnetic transition with increasing temperature, similar to that in the bulk (~369 K) [61]. The transition temperature ($T_C$) decreases with decreasing film thickness, from 330 K for 60 u.c. to 260 K for 7 u.c. However, below 6 u.c., the films exhibit insulating behavior in all measured $T$-range, thus identifying a thickness-dependent MIT at critical thickness $n_{cr}$ = 6 u.c.

### III A: Thick Films

Figure 2 displays the STEM/EELS structural and chemical compositional data obtained along two different sample orientations [see Fig. 1(a)], [100] on the left and [1-10] on the right, for a 40 u.c. film. Figure 2 (a) displays HAADF-STEM images taken along the [100] direction with the integrated EELS elemental profiles superimposed, and 2(b) shows the corresponding intensity profiles of different atomic sites. The results clearly show the position of the interface and the continuity of the perovskite stacking sequence across the interface without any dislocation. Figure 2(c & d) display the Mn and Ti, La and Sr distribution as a function of distance (in u.c.) from the interface, respectively. The interface between STO and LSMO is defined between the $TiO_2$ termination layer and the (La/Sr)O layer of LSMO [See Fig. 1(a)], so that its corresponding coordinate is -0.5 u.c. in horizontal axis (the same hereafter). There is no more than one u.c. intermixing of Mn/Ti and La/Sr, where the Sr concentration in the first (La/Sr)O layer is ~ 60% ($x$ ~ 0.6). In the [1-10] direction, the HAADF image of Fig 2(e) and La/Sr profiles of Fig. 2(i) present the identical conclusions of the [100] direction shown in Fig. 2 (a) and (d).

The quantitative analysis of the lattice constants and local displacements using STEM are summarized in Table I. The lattice spacing along the in-plane (IP) and OOP directions were determined from positions of A-site cations, by averaging 80 u.c. parallel to the interface. Figure 2(g) shows the OOP lattice constant as a function of distance from the interface obtained from Fig. 2(a) and (e). The blue and red solid lines mark the bulk lattice constant of STO (3.905 Å) and LSMO (3.87 Å), respectively. As shown in Fig. 2(g), the LSMO OOP $c$-axis lattice constant matches the bulk value except for the first few layers from the interface exhibiting slightly larger values than the bulk while still within the experimental error bar. Away from the interface, the lattice constant is slightly smaller than the bulk value (3.87 Å) determined by XRD because of the tensile strain induced by STO.

Table I: Structural data for 40 u.c LSMO films from STEM.

|  | Bulk (LSMO) | LSMO/STO |
|---|---|---|
| Tilt Angle $\alpha$ (deg.) | 6.80 | $5.80 \pm 0.09$ |
| In-plane ($a = b$) (Å) | 3.87 | $3.90 \pm 0.03$ |
| $c$-axis (Å) | 3.87 | $3.85 \pm 0.02$ |
| Mn-O(2) (Å) | 1.94 | $1.94 \pm 0.06$ |
| O(1)-O(2) (Å) | 2.74 | $2.76 \pm 0.06$ |
| Octahedron Volume (Å$^3$) | 9.67 | $9.90 \pm 0.17$ |

The MnO$_6$ octahedral tilt distortion in LSMO [see Fig. 1(g)] was determined from the oxygen positions in the ABF images along the [1-10] direction. The ABF-STEM image [Fig. 2(f)] shows a zig-zag pattern of oxygen atoms which is visible in the film (see the inset). The tilt angle, located at the $i^{th}$-u.c. from the interface, was calculated by $\alpha_i = \tan^{-1}\left(\Delta y_i / \Delta x_i\right)$ where $\Delta x_i$ and $\Delta y_i$ is the distance of neighboring oxygen along the $x$ ([110]) and $y$ ([001]) directions, respectively. The evolution of the tilt angle across the interface was obtained by averaging alternately over 50 u.c. parallel to the interface. The results are shown in Fig. 2(h). The substrate is in cubic structure without tilt of octahedral. A reduced OOT angle, $(3.5 \pm 1.3)°$ is clearly seen in the first u.c. of LSMO as compared to the bulk value of $(6.8°)$ [62]. Away from the interface, the tilt angle gradually increases to an averaged value of $(5.80 \pm 0.09)°$ in the film, which is still smaller than the bulk value. Such a reduction of tilt angle can be attributed to effects of substrate-induced tensile strain in the film, suppressing the tilt distortion of the MnO$_6$ octahedra. The MnO$_6$ IP rotation is difficult to identify due to the overlap of O atoms. Since all the films are fully strained by the substrate but still display excellent coherent structure across the interface, we do not expect any obvious IP MnO$_6$ rotation. Based on the STEM and XRD results, the detailed structures of the LSMO film are presented in Table I. From structural point of view, the reduced tilt of first few unite cell should enhance the metallicity based on the double-exchange mechanism. Therefore, the insulating behavior below 6 u.c. cannot be mainly caused by the substrate-induced strain.

The composition near the surface was determined by *in-situ* ARXPS. Figure 3(a) presents the ARXPS spectra of Mn 2p, Sr 3d and La 4d peaks as a function of emission angle ($\theta$) for a 65 u.c. LSMO film. The finite inelastic mean free path of photoelectrons in ARXPS provides chemical composition information at different depths by varying the emission angle ($\theta$). The angle dependence of the relative core-level intensity ratios, $I_{Sr3d}/I_{La4d}$ for 20, 40, and 65 u.c. (thick films) LSMO films are shown in Fig. 3(b). The ratio $I_{Sr3d}/I_{La4d}$ for these three different thicknesses have identical angular dependence, indicating universal variation of Sr concentration near the surface of thick films. These measurements cannot be interpreted for the films with thickness < 15 u.c. because of the considerable Sr contributions from the STO substrate.

A qualitative inspection of Fig. 3(a) yields several important observations. First, the intensity of the Mn 2p compared to either the Sr 3d or La 4d decreases dramatically with increasing emission angle. This indicates that the surface is (La/Sr)O terminated which will be discussed later. Secondly, the intensity of the Sr 3d compared to the La 4d increases as the emission angle increase, indicating a Sr rich surface found with the fitting procedure described presently.

To obtain the Sr vs. La concentration, the data is fitted to the model function for the intensity ratio $R_{AB}(\theta)$ between two elements (atom A and B) [31], given by:

$$R_{AB}(\theta) = \frac{I_A}{I_B} = \frac{\sigma_A \cdot T_{rA} \cdot \sum_i f_i^A \cdot \exp\left(\frac{-id}{\lambda_A \cos\theta}\right)}{\sigma_B \cdot T_{rB} \cdot \sum_j f_j^B \cdot \exp\left(\frac{-jd}{\lambda_B \cos\theta}\right)} \quad \text{------ (1)},$$

where $\sigma$ is the photoionization cross section of each element obtained the library database from Spec software [63], $T_r$ is the transmission coefficient of the analyzer, varying as the kinetic energy of the emitted electrons [64], $d$ the interlayer spacing, $\lambda$ the inelastic mean free path of the photoelectrons calculated based on the Tanuma, Powell and Penn algorithm (TPP2M method) [65], $\theta$ the emission angle with respect to the surface normal, and $f_i$ the atomic fraction of element (A or B, which are Sr and La, respectively) at the $i^{th}$-layer, which is assumed to have an exponential segregation profile [26]:

$$f = b + \delta_s \exp(-id/l_s) \quad \text{------ (2)},$$

where $b$ is the bulk fraction of an element (1/3 for Sr and 2/3 for La in our case), and $\delta_s$ and $l_s$ are two parameters determined by fitting, which present the deviation of concentration from the bulk value and depth of segregation, respectively. Cation segregation of transition-metal oxides impacts the reactions that are often critical to the overall device performance and applications [66-74]. Sr segregation is commonly observed in LSMO, which cannot be avoided by changing growth condition or any post anneal treatment. For LSMO films, the insulating surface is a polar surface with ($La^{3+}$/$Sr^{2+}$)$O^{2-}$ layer termination. During the initial growth, and when the film is still not metallic, increasing Sr concentration at the surface helps to neutralize the polar surface by reducing the surface energy, although other factors such as surface reconstruction need to be considered. In many oxide surfaces, reducing surface polarity and/or interface polar discontinuity is the primary driving force for the change of surface/interface composition and lattice structure.

The Sr concentration profiles are fitted using an exponential function given in Eq. (2) for both surface and interface regions by replacing d by n-d for the interface. The fitting results are given in Table II. The Sr profiles near the surface of 20, 40, and 65 u.c. films are identical for the all film thicknesses, as shown in the right side of Fig. 3(c). The Sr concentration of the top layer reaches ~ 0.6 (increased by ~ 80%) and the deviation from the bulk value extends to more than 3 u.c. from the surface. The results are inconsistent with the previous rock-salt structured model with the very top SrO layer [34]. With the rock-salt structured model, the fitting result of the intensity ratio between Sr and La cores would be 1.66 at the photoelectron emission angle $\theta = 81°$, which is about twice value of measured $I_{sr3d}/I_{La4d}$ shown in the Fig. 3(b).

Table II. Fitting results for Sr profiles near the surface and interface of LSMO films

|  | Thickness $n$ (u.c.) | $l$ (u.c.) | δ |
|---|---|---|---|
| Surface (ARXPS results) | 20, 40 & 65 | 1.02 ± 0.16 | 0.24 ± 0.01 |
| Interface (STEM results) | 40 | 0.26 ± 0.06 | 0.22 ± 0.01 |
| | 4 & 8 | 0.95 ± 0.18 | 0.31 ± 0.02 |

To determine the interface chemical composition, the layer-by-layer composition profile of films were also determined by analyzing the STEM EELS spectra [see Fig. 2(a)]. The La $M_{4,5}$, Sr $L_{2,3}$, Mn $L_{2,3}$ and Ti $L_{2,3}$ edges of the EELS spectra were integrated for each column and then added across 6 u.c. line profiles parallel to the interface to provide layer-by-layer elemental profiles. To improve the statistics, we have taken STEM/EELS data from three different scanning areas and then averaged. To quantify the amount of Sr segregation, a calibration was achieved by referring the EELS intensity profiles of Sr acquired from the middle range (10 - 30 u.c.) of 40 u.c. film as the defined 33% Sr concentration in LSMO. This is justified by the fact that the 40 u.c. LSMO film shows almost the same transport and magnetic properties as the bulk LSMO with nominal Sr doping $x = 1/3$. An additional calibration was performed by measuring the La/Sr intensity ratio in bulk LSMO.

By combining STEM/EELS with ARXPS results, the layer-by-layer Sr concentrations near the interface and surface for LSMO thick films were obtained and are shown in Fig. 3(c). The only detailed STEM measurement was performed on a 40 u.c. sample, but we have no reason to expect different profiles for other thicknesses. Therefore, we conclude that for all LSMO films thicker than 20 u.c., there is Sr segregation at the surface and little if any intermixing at the interface [75].

The STEM images and EELS profiles near the surface of the LSMO films were also measured and confirmed the results from our ARXPS measurements. In general, STEM cannot be used to determine the Sr segregation to the free surface, since damage to the surface could occur in *ex-situ* STEM sample preparation. To avoid such uncertainty, we have characterized a 40 u.c. LSMO film capped with amorphous $BaTiO_3$ by performing STEM imaging and EELS spectroscopy mapping. We found Sr surface segregation similar to what was concluded from ARXPS measurements. In addition, we observed that the thick film surface is terminated with a (La/Sr)O layer.

The inhomogeneous Sr concentration in thick LSMO films suggests that there may be a systematic variation pattern in the transport properties with film thickness. Figure 4 (a) presents the thickness dependence of conductivity $\sigma$ measured at $T = 4$ K for all films above the critical thickness ($n_{cr} = 6$ u.c.). For a uniform metallic film, if each layer of the film had the same metallic behavior, one would expect to have the same conductivity value at a fixed temperature regardless film thickness. However, this is not the case here. As shown in Fig. 4(a), the measured conductivity decreases nonlinearly with decreasing film thickness ($n$). Interestingly, the product of conductivity and thickness, $\sigma \cdot n$, linearly depends on thickness $n$. As shown in Fig. 4(a), $\sigma \cdot n$ vs. $n$ for all the data shown in Fig. 5(a) can be fitted by a linear function $\sigma \cdot n = \sigma_b \cdot (n - n_0)$ with the 'dead' layer thickness $n_0 = 6.5 \pm 0.9$ u.c. and the conductivity $\sigma_b = 7344.6 \pm 176.5$ $\Omega^{-1} \cdot cm^{-1}$ of inside metallic part of films. Interestingly, the value of $\sigma_b$ and $n_0$ are similar to the bulk value of conductivity $\sim 10^4$ $\Omega^{-1} \cdot cm^{-1}$ [61] and the critical thickness ($n_{cr}$) of the film [58], respectively. If the films were uniform, the linear function fitting would give $n_0 = 0$. This implies that, if we exclude a certain thickness ($n_0$) of nonmetallic layers regardless of film thickness, the film exhibits a thickness-independent bulk-like conductivity ($\sigma_b$) or becomes uniform. All the films show a Sr-rich surface and interface, thus suggesting that these nonmetallic layers should exist near the surface and interface. During the transport measurement, when these layers are much less conducting than the inside part of the film, the current will go through the inside part of films and "ignore" the existence of these nonmetallic layers, resulting in the offset of the linear fitting shown in Fig. 4(a).

Figure 4(b) depicts a simple picture based on the conductivity measurements and layer-by-layer composition characterization, with the yellow shaded off-stoichiometry regions at the interface and surface exhibiting nonmetallic behavior and the blue region in the center part being the bulk-like metallic LSMO. The similarity between the value of $n_0$ and $n_{cr}$ indicates that as the nonmetal layers at the surface overlap the nonmetallic layer(s) at interface (as $n \rightarrow n_0$ or $n_{cr}$) the film becomes completely insulating, Thus, the off-stoichiometry is the primary cause of dead layer. On the other hand, this is an oversimplified model since the conductivity can vary layer by layer away from the interface and surface. Furthermore, as depicted in Fig. 4(b), the nonmetallic region near the interface and surface is bigger than the off-stoichiometry region measured from these thick films, implying that such deviation of chemical composition and/or structure in the ultrathin film case may be more severe than what we observed in the thick films. The contribution to the formation of the dead layer also can be oxygen vacancy and strain. On the other hand, oxygen vacancy as a primary contribution can be ruled out by our experiment. Increasing oxygen pressure for the film growth or any post annealing in oxygen does not further reduce the dead layer thickness. Strain effect should not be the main contribution for the dead layer either since the reduction of tilt angle observed near the interface [Fig. 2(h)] would enhance the film metallicity based on the double-exchange mechanism.

### III.B, Thin Films

As we have discussed above and depicted in Fig. 4, the existence of nonmetallic and off-stoichiometric layers near the surface and interface persists for all thick films and is responsible for the dead layer behavior. This implies that the chemical profiles near the surface and interface would remain the same with further reduction of film thickness. Interestingly, this is not the case based on our study of ultrathin films. When film thickness is less than ~10 u.c., the effective role of surface and interface becomes more pronounced which leads to the enhanced deviation of chemical composition (increased Sr) in the films. The HAADF-STEM images and EELS profiles in Fig.5 (a) and (b) for uncapped 8 and 4 u.c. LSMO films show a roughly 2 u.c. intermixing region at the interface, compared to the only one u.c. intermixing interface region in thick films. The averaged intensity profiles of Fig. 5 (c) and (d) further indicate the difference in thin films from thick films (presented in Fig. 2(b)). To quantitatively exhibit the change in thin films, we have used EELS to determine the layer-by-layer concentrations of Mn, Ti, La and Sr near the interface and surface. The results are displayed in Fig. 5 (e), (g), (f), and (h) for 8 and 4 u.c. films, respectively. The Mn and Ti profile [see Fig. 5(e) and (f)] at the interface confirms an over 2.u.c intermixing region at the interface. The La/Sr profiles [see Fig. 5(h) and (h)] also show a much broader intermixing region for thin films. Although the surface results contain large error bars due to the weaker intensity and possible surface damage during TEM sample processing, the profiles clearly demonstrate Sr segregation on the thin film surface in Fig. 5 (g) and (h).

We have also studied the LSMO films *in-situ* capped with amorphous BTO grown at room temperature. Compared with the uncapped thin films, similar behavior can be concluded as shown in Fig. 5(i-l). The only difference here is that due to the existence of the BTO capping layer, there is a small amount of Ba atoms diffusing to the surface layer of LSMO. To summarize, the Sr segregation in both thick and thin films is presented in the Fig. 5 (m). For thin films, Sr concentration at the first layer (0 u.c.) at the interface is about 0.64, which is comparable to the thick film result. However, at the second layer (1 u.c.), Sr concentration is higher for the thin films (~ 0.42) than the thick films (~ 0.34).

### III.C, Surface Termination

The surface termination of LSMO films exhibits a thickness-dependent behavior, evolving from mixed $MnO_2$- and (La/Sr)O-layers to eventual (La/Sr)O-layer termination with increasing film thickness, as determined by ARXPS. To avoid the effect of the substrate Sr contribution on Sr-core spectra, we used the ratio of the La 4d to the Mn 2p core levels to extract the information of the surface termination. The ratios as a function of emission-angle for different thickness of films are displayed in Fig. 6(a). For films thicker than 20 u.c. the core intensity ratios of La/Mn core have identical angular dependence within the error bar.

The increase of La/Mn ratio with increasing emission angle indicates that the surface is terminated by a (La/Sr)O layer. In contrast, the ratio for thin films (4, 6, and 10 u.c.) exhibit similarly small angle-dependence compared to that for the thick films (see Fig. 6a). Moreover, the ratio has a thickness-dependent offset, increasing with increasing thickness. Three scenarios of the surface could explain these data for thin films; either a mixed surface termination between $MnO_2$- and (La/Sr)O-layer without any Sr surface segregation, a pure (La/Sr)O layer termination with changing Sr surface segregation with thickness, or a mixed surface termination with thickness-dependent Sr surface segregation might explain this behavior. However, the first two scenarios are unlikely. STEM/EELS observations of surface segregation (see Fig. 5) rules out the first scenario. In addition, as there is an extra (La/Sr)O layer at the initial deposition on the $TiO_2$-terminated substrate, one would expect a decrease in the ratio of La/Mn at normal emission ($\theta$) with the initial increase of film thickness, which contradicts the data shown in Fig. 6(b). The second scenario is also unlikely when comparing the simulation with experimental results at $\theta = 0°$ and $81°$ shown in Fig. 6(b). If we assume that the (La/Sr)O layer is the termination layer and use the Sr surface segregation profile determined from thick films (see Table II) as an approximation, the La/Mn core intensity ratio is calculated and shown as the dashed curves in Fig. 6(b). There is a clear deviation of calculated ratio from the experimental data in the thin film region, especially for the data taken at the emission of $81°$. Therefore, the surface has both mixed termination and thickness-dependent Sr segregation for thin films.

To further quantify the evolution of surface termination with film thickness, we have analyzed the data for the thickness dependence of the La/Mn ratio at $0°$ and $81°$ shown in Fig 6(b) by fitting to the model function for the intensity ratio. Assuming the surface is composed with a fraction ($y$) of (La/Sr)O layer and (1-$y$) of $MnO_2$ layer for a film with given thickness, the intensity ratio $R_{AB}(\theta)$ given in Eq. (1) can be modified for the La/Mn ratio and given in Eq. (3).

$$R_{La/Mn} = \frac{\sigma_{La} T_{rLa}}{\sigma_{Mn} T_{rMn}} \frac{\left[ y \sum_{i=0}^{n} f_i^{La} \exp\left( \frac{-id}{\lambda_{La} \cdot \cos\theta} \right) + (1-y) \sum_{m=0}^{n-1} f_m^{La} \exp\left( \frac{-(m+0.5)d}{\lambda_{La} \cdot \cos\theta} \right) \right]}{\left[ y \sum_{j=0}^{n} \exp\left( \frac{-(j+0.5)d}{\lambda_{Mn} \cdot \cos\theta} \right) + (1-y) \sum_{l=0}^{n-1} \exp\left( \frac{-ld}{\lambda_{Mn} \cdot \cos\theta} \right) \right]} \quad \text{------ (3)}$$

where $f_i$ ($f_m$) is the atomic fraction of element at the $i^{th}$ ($m^{th}$) layer. By using this model, the results are presented in Fig. 6(b). The (La/Sr)O termination fraction ($y$) as a function of film thickness is obtained and displayed in the inset of Fig. 6(b). The mixed surface with dominant $MnO_2$ layer termination occurs for the 4 u.c. film but the fraction of (La/Sr)O-layer termination increases with the film thickness. Eventually, the surface is completely terminated with a (La/Sr)O-layer. Although Sr segregation helps to reduce the surface energy, qualitatively, breaking $MnO_6$ to have a $MnO_2$ termination could cost more energy than (La/Sr)O-termination [76]. The evolution of surface termination observed here indicates that (La/Sr)O is the energetically favorable termination layer for the surface of LSMO films. The mixed surface termination in the thin film can be understood as a substrate effect during the initial growth because the substrate is terminated with a $TiO_2$-layer.

The STEM/EELS results further confirm such thickness dependent evolution of the surface termination. To examine the pristine surface of LSMO films, we grew both 40 u.c. and 6 u.c. LSMO films under the same conditions but *in-situ* capped them with a layer of amorphous BTO deposited at room temperature and performed STEM/EELS measurements. As shown in Fig. 6(c) and (d), a clear difference is demonstrated between these two surfaces. For the 40 u.c. LSMO film, the surface is uniformly terminated by (La/Sr)O-layer seen from both HAADF images and elemental-specific EELS spectroscopic mapping. However, for the 6 u.c. LSMO film, mixed termination is seen in the image. Especially, intermixing between Mn and Ti or La and Ba appears in the EELS mapping, confirming the termination difference between the thin and thick films.

## IV. Summary

We demonstrate that metallic ferromagnetic LSMO films grown on $SrTiO_3$ substrates exhibit Sr segregation at the surface and single layer or less intermixing at the interface. This property is universal for LSMO films of 20 u.c. thickness or more. Below 20 u.c., the surface of the films appears mixed (La/Sr)O with $MnO_2$ termination, imitating the substrate $TiO_2$ termination at the interface. Detailed measurements of the transport properties as a function of film thickness indicate that there are inherent nonmetallic layers independent of film thickness, driving the film insulating at a critical thickness.

We would like to thank Jing Tao for the assistance of the STEM experiments. This work was primarily supported by U.S. DOE under Grant No. DOE DE-SC0002136. Y. Z is supported by the Materials Science and Engineering Division, DOE-BES, under Contract No: DE-AC02-98CH10886. This research used resources of the Center for Functional Nanomaterials, which is a U.S. DOE Office of Science Facility, at Brookhaven National Laboratory under Contract No. DE-SC0012704.

**References:**

1. J. E. Demuth, B. N. J. Persson, and A.J. Schell-Sorokin, *Phys. Rev. Lett.* **51**, 2214 (1983).

2. W. Meevasana, P.D.C. King, R.H. He, S-K. Mo, M. Hashimoto, A. Tamai, P. Songsiriritthigul, F. Baumberger, and Z.-X. Shen, *Nat. Mater.* **10**, 114 (2011).

3. R.G. Moore, Jiandi Zhang, V.B. Nascimento, R. Jin, J. Guo, G.T. Wang, Z. Fang, D. Mandrus, and E.W. Plummer, *Science* **318**, 615 (2007).

4. R. G. Moore, V.B. Nascimento, Jiandi Zhang, J. Rundgren, R. Jin, D. Mandrus, and E.W. Plummer, *Phys. Rev. Lett.* **100**, 066102 (2008).

5. C. Chen, J. Kim, V.B. Nascimento, Z. Diao, J. Teng, B. Hu, G. Li, F. Liu, Jiandi Zhang, R. Jin, and E.W. Plummer, *Phys. Rev.* B **95**, 085420 (2016).

6. A. Ohtomo and H. Y. Hwang, *Nature* **427**, 423 (2004).

7. A. Brinkman, M. Huijben, M. van Zalk, J. Huijben, U. Zeitler, J. C. Maan, W.G. van der Wiel, G. Rijnders, D.H.A. Blank, and H. Hilgenkamp, *Nat. Mater* **6**, 493 (2007).

8. N. Reyren, S. Thiel, A.D. Caviglia, L.F. Kourkoutis, G. Hammerl, C. Richter, C.W. Schneider, T. Kopp, A.-S. Rüetschi, D. Jaccard, M. Gabay, D.A. Muller, J.-M. Triscone, and J. Mannhart, *Science* **317**, 1196 (2007).

9. J.A. Bert, B. Kalisky, C. Bell, M. Kim, Y. Hikita, H.Y. Hwang, and K. A. Moler, *Nat. Phys.* **7**, 767 (2011).

10. N. Nakagawa, H. Y. Hwang, and D. A. Muller, *Nat. Mater.* **5**, 204 (2006).

11. P. R. Willmott, S. A. Pauli, R. Herger, C. M. Schlepütz, D. Martoccia, B. D. Patterson, B. Delley, R. Clarke, D. Kumah, C. Cionca, and Y. Yacoby, *Phys. Rev. Lett.* **99**, 155502 (2007).

12. A.S. Kalabukhov, Y.A. Boikov, I.T. Serenkov, V.I. Sakharov, V.N. Popok, R. Gunnarsson, J. Börjesson, N. Ljustina, E. Olsson, D. Winkler, and T. Claeson, *Phys. Rev. Lett.* **103**, 146101 (2009).

13. W. Siemons, G. Koster, H. Yamamoto, W.A. Harrison, G. Lucovsky, T.H. Geballe, D.H.A. Blank, and M.R. Beasley, *Phys. Rev. Lett.* **98**, 196802 (2007).

14. G. Herranz, M. Basletić, M. Bibes, C. Carrétéro, E. Tafra, E. Jacquet, K. Bouzehouane, C. Deranlot, A. Hamzić, J.-M. Broto, A. Barthélémy, and A. Fert, *Phys. Rev. Lett.* **98**, 216803 (2007).

15. R. Pentcheva and W. E. Pickett, *Phys. Rev. Lett.* **102**, 107602 (2009).

16. K. Yoshimatsu, T. Okabe, H. Kumigashira, S. Okamoto, S. Aizaki, A. Fujimori, and M. Oshima, *Phys. Rev. Lett.* **104**, 147601 (2010).

17. J. Xia, W. Siemons, G. Koster, M.R. Beasley, and A. Kapitulnik, *Phys. Rev. B* **79**, 140407R (2009).

18. A.V. Boris, Y. Matiks, E. Benckiser, A. Frano, P. Popovich, V. Hinkov, P. Wochner, M. Castro-Colin, E. Detemple, V.K. Malik, C. Bernhard, T. Prokscha, A. Suter, Z. Salman, E. Morenzoni, G. Cristiani, H.-U. Habermeier, and B. Keimer, *Science* **332**, 937 (2011).

19. R. Scherwitzl, S. Gariglio, M. Gabay, P. Zubko, M. Gibert, and J.-M. Triscone, *Phys. Rev. Lett.* **106**, 246403 (2011).

20. M. Izumi, Y. Konishi, T. Nishihara, S. Hayashi, M. Shinohara, M. Kawasaki, and Y. Tokura, *App. Phys. Lett.* **73**, 2497 (1998).

21. M. Huijben, L.W. Martin, Y.H. Chu, M.B. Holcomb, P. Yu, G. Rijnders, D.H.A. Blank, and R. Ramesh, *Phys. Rev.* B **78**, 094413 (2008).

22. Z. Zhong, M. Wallerberger, J.M. Tomczak, C. Taranto, and N. Parragh, *Phys. Rev. Lett.* **114**, 246401 (2015).

23. A. Tebano, C. Aruta, S. Sanna, P.G. Medaglia, G. Balestrino, A.A. Sidorenko, R. De Renzi, G. Ghiringhelli, L. Braicovich, V. Bisogni, and N.B. Brookes, *Phys. Rev. Lett.* **100**, 137401 (2008).

24. V. Bisogni, S. Catalano, R.J. Green, M. Gibert, R. Scherwitzl, Y. Huang, V.N. Strocov, P. Zubko, S. Balandeh, J.-M. Triscone, G. Sawatzky, and T. Schmitt, *Nat. Commun.* **7**, 13017 (2016).

25. M.B. Lepetit, B. Mercey, and C. Simon, *Phys. Rev. Lett.* **108**, 087202 (2012).

26. H. Dulli, P. Dowben, S.H. Liou, and E.W. Plummer, *Phys. Rev.* B **62**, R14629 (2000).

27. T.T. Fister, D.D. Fong, J.A. Eastman, P.M. Baldo, M.J. Highland, P.H. Fuoss, K.R. Balasubramaniam, J.C. Meador, and P.A. Salvador, *Appl. Phys. Lett.* **93**, 151904 (2008).

28. R. Bertacco, J.P. Contour, A. Barthemy, and J. Olivier, *Surf. Sci.* **511**, 366 (2002).

29. H. Jalili, J.W. Han, Y. Kuru, Z. Cai, and B. Yildiz, *J. Phys. Chem. Lett.* **2**, 801 (2011).

30. K. Katsiev, B. Yildiz, K. Balasubramaniam, and P.A. Salvador, *Appl. Phys. Lett.* **95**, 092106 (2009).

31. F. Monsen, F. Song, Z.S. Li, J.E. Boschker, T. Tybell, E. Wahlström, and J.W. Wells, *Surf.*

*Sci*. **606**, 1360 (2012).

32. L. Poggini, S. Ninova, P. Graziosi, M. Mannini, V. Lanzilotto, B. Cortigiani, L. Malavolti, F. Borgatti, U. Bardi, F. Totti, I. Bergenti, V.A. Dediu, and R. Sessoli, *J. Phys. Chem.* C **118**, 13631 (2014).
33. P. Decorse, E. Quenneville, S. Poulin, M. Meunier, and A. Yelon, *J. Vac. Sci. Technol.* A **19**, 910 (2001).
34. Z. Li, Michel Bosman, Zhen Yang, Peng Ren, Lan Wang, Liang Cao, Xiaojiang Yu, Chang Ke, M. B.H. Breese, A. Rusydi, W. Zhu, Z. Dong, and Yong Lim Foo, *Adv. Funct. Mater.* **22**, 4312(2012).
35. G. Agricola, *De Re Metallica* (J. Froben and N. Episcopius, Basel, 1556, translated by H.C. Hoover and L.H. Hoover, London, 1912, reprinted by Dover Publications, New York, 1950).
36. G. Gottstein, and L. S. Shvindlerman, *Grain boundary migration in metals thermodynamics, kinetics, Applications,* CRC Press (1999).
37. A.D. Rollet, G. Gottstein, L.S. Shvindlerman, and D.A. Molodov, *Zeitschrift für Metallkunde* **95**, 226 (2004).
38. X. Yin, Muhammad Aziz Majidi, X. Chi, Peng Ren, Lu You, Natalia Palina, Xiaojiang Yu, Caozheng Diao, Daniel Schmidt, Baomin Wang, Ping Yang, Mark B H Breese, Junling Wang and Andrivo Rusydi, *NPG Asia Materials* **7**, e196 (2015).
39. F.Y. Bruno, J. Garcia-Barriocanal, M. Varela, N. M. Nemes, P. Thakur, J. C. Cezar, N. B. Brookes, A. Rivera-Calzada, M. Garcia-Hernandez, C. Leon, S. Okamoto, S. J. Pennycook, and J. Santamaria, *Phys. Rev. Lett.* **106**, 147205 (2011).
40. O. Shapoval, S. Huehn, J. Verbeeck, M. Jungbauer, A. Belenchuk, and V. Moshnyaga, *J. Appl. Phys*. **113**, 17C711 (2013).
41. W. Lee, J. W. Han, Y. Chen, Z. Cai, and B. Yildiz, *J. Am. Chem. Soc.* **135**, 7909 (2013).
42. H. Boschker, J. Verbeeck, R. Egoavil, S. Bals, G. van Tendeloo, M. Huijben, E.P. Houwman, G. Koster, D.H.A. Blank, and G. Rijnders, *Adv. Funct. Mater.* **22** 2235 (2012).
43. P. Wynblatt, R.C. Ku, in *Interfacial Segregation* (Eds. W.C. Johnson, J.M. Blakely), American Society for Metals, Metals Park (1979).
44. *Surfaces and Interfaces of Ceramic Materials*, Eds. L.C. Dufour and C. Monty, Springer (1989).
45. K. Verbist, O. I. Lebedev, G. Van Tendeloo, M. A. J. Verhoeven, A. J. H. M. Rijnders, D.

H. A. Blank, and H. Rogalla, *Appl. Phys. Lett.* **70**,1167 (1997).

46. K. L. Hoyer, A. H. Hubmann, and A. Klein, *Phys. Status Solidi* A **214**, 1600486 (2017).

47. Y. Shigesato, S. Takaki, and T. Haranoh, J. Appl. Phys. **71**, 3356 (1992).

48. O. Brandt, L. Tapfer, and K. Ploog, *Appl. Phys. Lett.* **61**, 2814 (1998).

49. R. Magri, and A. Zunger, *Phys. Rev.* B **64**, 081305(R) (2001).

50. C. Li, L. Wang, W. Chen, L. Lu, H. Nan, D. Wang, Y. Zhang, Y. Yang, and C. Jia, *Adv. Mater. Interfaces* **5**, 1700972 (2018).

51. W. T. Geng, A. J. Freeman, and R. Q. Wu, *Phys. Rev.* B **63**, 064427 (2001).

52. M. Všianská and M. Šob, *Phys. Rev.* B **84**, 014418 (2011).

53. R. Herger, P. R. Willmott, C. M. Schlepütz, M. Björck, S. A. Pauli, D. Martoccia, B. D. Patterson, D. Kumah, R. Clarke, Y. Yacoby, and M. Döbeli, *Phys. Rev. B* **77**, 085401 (2008).

54. L. Jin, C. Jia, I. Lindfors-Vrejoiu, X. Zhong, H. Du, and R. E. Dunin-Borkowski, *Adv. Mater. Interfaces* **3**, 1600414 (2016)

55. F. Song, Å. F. Monsen, Z. S. Li, J. W. Wells, and E. Wahlström, *Surf. Interface Anal.* **45**, 1144(2013).

56. A. Ohtomo, D.A. Muller, J.L. Grazul and H.Y. Hwang, *Nature* **419**, 378 (2002).

57. H.W. Guo, Z. Wang, S. Dong, S. Ghosh, M. Saghayezhian, L. Chen, Y. Weng, A. Herklotz, T.Z. Ward, R. Jin, S.T. Pantelides, Y. Zhu, Jiandi Zhang and E.W. Plummer, *PNAS* **114**, E5092 (2017)

58. Z. Liao, F. Li, P. Gao, L. Li, J. Guo, X. Pan, R. Jin, E.W. Plummer, and Jiandi Zhang, *Phys. Rev.* B **92**, 125123 (2015).

59. S. J. Pennycook, "Scanning Transmission Electron Microscopy: Z-Contrast Imaging", *Characterization of Materials*. 1–29 (2012).

60. V.I. Nefedov and O.A. Baschenko, *J. Electron Specctrosc. Relat. Phenom.* **47**, 1 (1988).

61. A. Urushibara, Y. Moritomo, T. Arima, A. Asamitsu, G. Kido, and Y. Tokura, *Phys. Rev.* B **51**, 14103 (1995).

62. M. C. Martin, G. Shirane, Y. Endoh, K. Hirota, Y. Moritomo, and Y. Tokura, *Phys. Rev*. B **53**, 14285 (1996).

63. http://www.specs.de/cms/front_content.php?idart=920

64. Technical Manual of PHI Model 10-360 Precision Energy Analyzer.

65. S. Tanuma, C. J. Powell, and D. R. Penn, *Surf. Interface Anal.* **21**,165 (1994)

66. Z. Cai, M. Kubicek, J. Fleig, and B. Yildiz, *Chem. Mater.* **24**, 1116 (2012).

67. S.J. Jiang, *Solid State Electrochem.* **11**, 93 (2007).

68. W. Jung, and H.L. Tuller, *Energy Environ. Sci.* **5**, 5370 (2012).

69. T.H. Shin, S. Ida, and T. Ishihara, *J. Am. Chem. Soc.* **133**, 19399 (2011).

70. C.-Y. Tsai, A.G. Dixon, Y.H. Ma, W.R. Moser, and M.R. Pascucci, *J. Am. Ceram. Soc.* **81**, 1437 (1998).

71. L. Yang, L. Tan, X. Gu, W. Jin, L. Zhang, and N. Xu, *Ind. Eng.Chem. Res.* **42**, 2299 (2003).

72. E. Levi, A. Mitelman, D. Aurbach, and M. Brunelli, *Chem. Mater.* **19**, 5131 (2007).

73. D. Serrate, J.M. De Teresa, J. Blasco, M.R. Ibarra, L. Morellon, and C. Ritter, *Appl. Phys. Lett.* **80**, 4573 (2002).

74. F. Stavale, X. Shao, N. Nilius, H.-J. Freund, S. Prada, L. Giordano, and G. Pacchioni, *J. Am. Chem. Soc.* **134**, 11380 (2012),

75. Hangwen Guo, Zhen Wang, Shuai Dong, Saurabh Ghoshd, Mohammad Saghayezhian, Lina Chen, Yakui Weng, Andreas Herklotz, Thomas Z Ward, Rongying Jin, Sokrates T. Pantelides, Yimei Zhu, Jiandi Zhang & E.W. Plummer, *PNAS* **114** (26), E5062 (2017).

76. M. M. Kuklja, E. A. Kotomin, R. Merkle, Yu. A. Mastrikov and J. Maier, *Phys. Chem. Chem. Phys*. **15**, 5443 (2013).

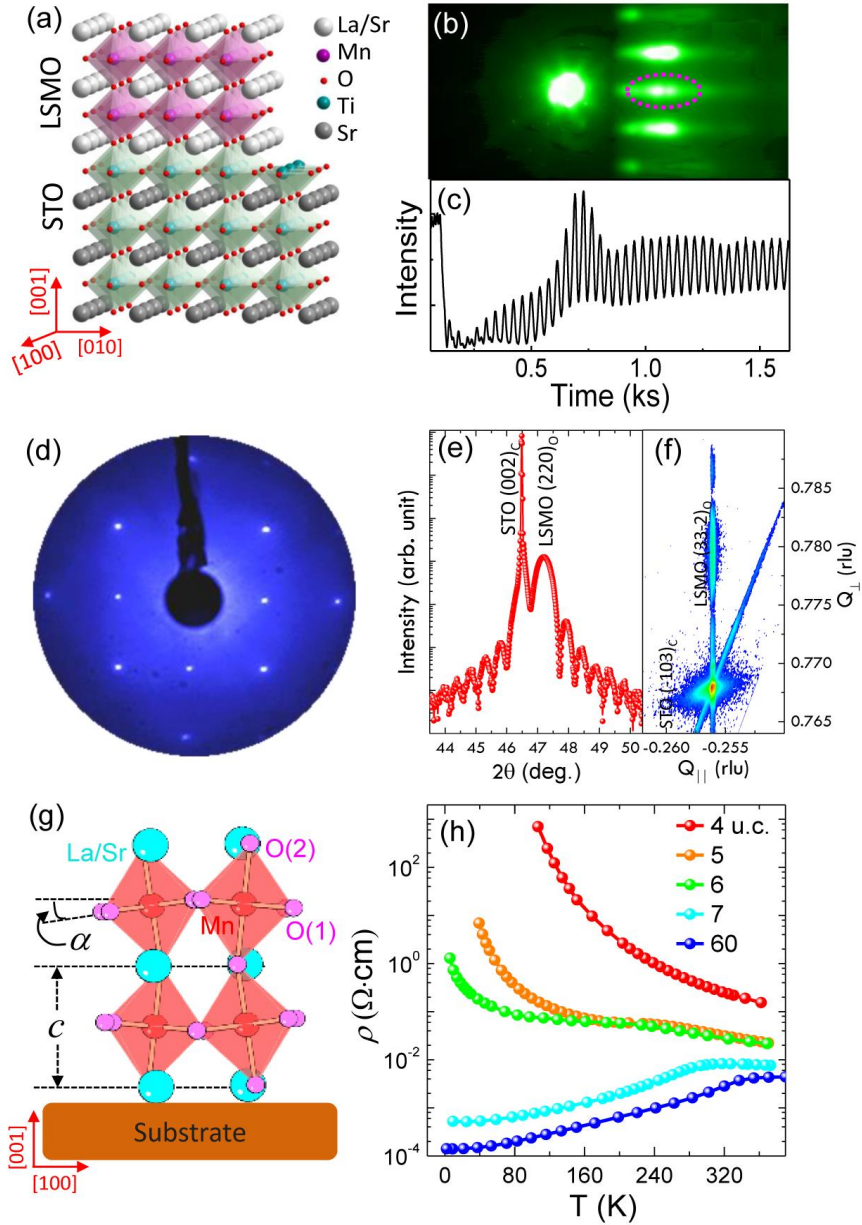

**Figure 1** (*Color online*) (a) Schematic lattice structure of LSMO/STO (001); RHEED (b) image and (c) intensity oscillation for the 40 u.c. LSMO film. (d) Typical LEED image from the surface of LSMO films. (e) Coupled symmetric ($\theta$-$2\theta$) XRD around substrate [002] spot. (f) Reciprocal lattice mapping around STO [-103], where $Q_\perp$ is measured in a direction perpendicular to the interface. (g) Schematic picture of the octahedron tilt in LSMO. (h) *T*- dependence of resistivity of LSMO films with different thickness.

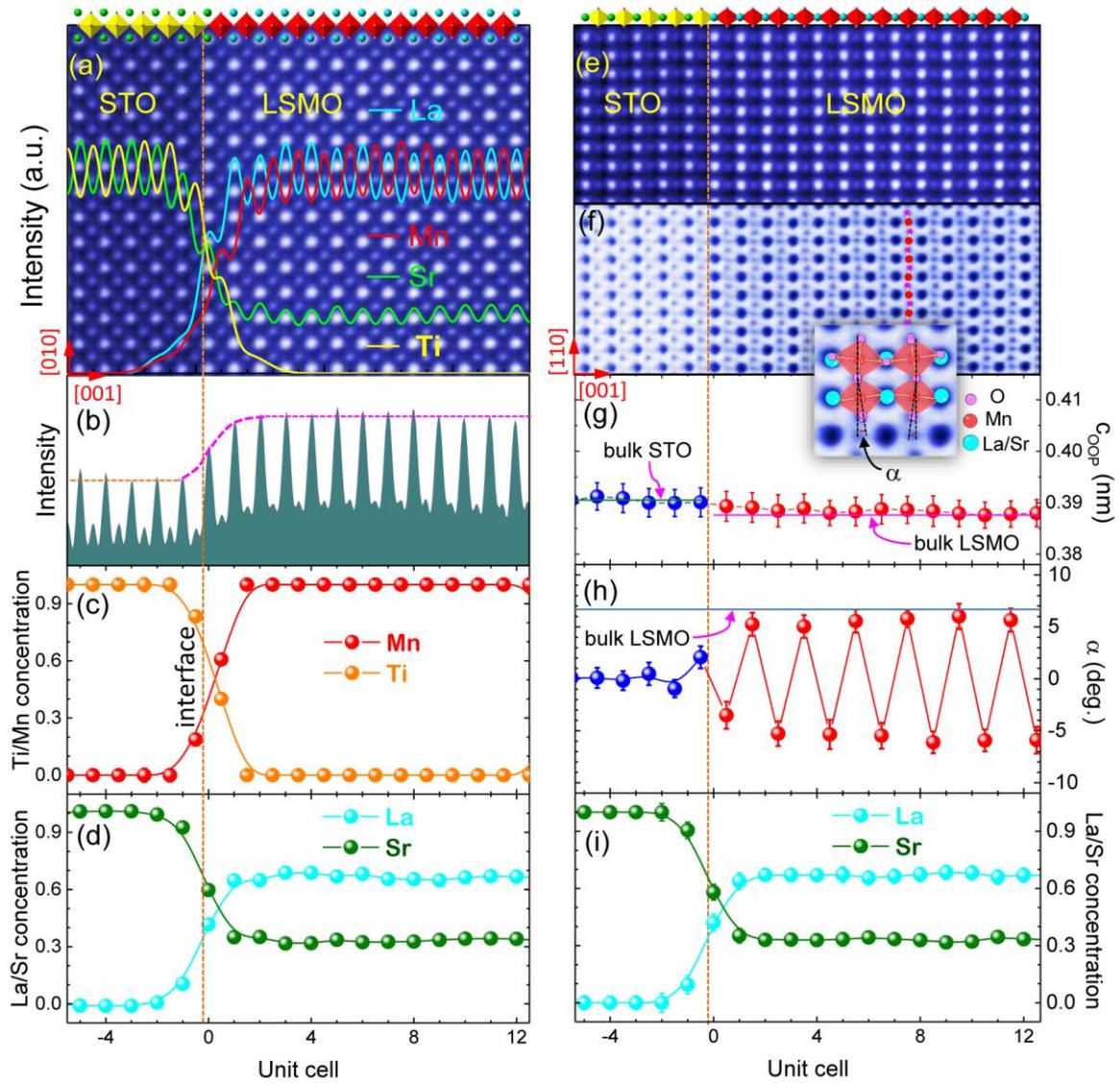

**Figure 2** (*Color online*) STEM data for 40 u.c. LSMO grown on STO (001) for two different sample alignments, [100] to the left and [1-10] to the right. (a) HAADF-STEM image overlapped with the integrated EELS elemental profiles for La, Sr, Ti and Mn across the LSMO/STO interface. (b) The intensity in the HAADF image across the interface. (c) Mn and Ti, and (d) La/Sr concentration profiles as a function of distance (unit cells) from the interface, respectively. The interface is marked by the dashed line. (e) HAADF- and (f) ABF-STEM image in [1-10] orientation, with inset showing distortion of the octahedral in the LSMO. (g) The measured c-axis lattice constant, (h) octahedral tilt angle and (i) La/Sr concentration profiles as function of distance from the interface, respectively, with the solid lines being the bulk value and the dashed line the XRD determined value.

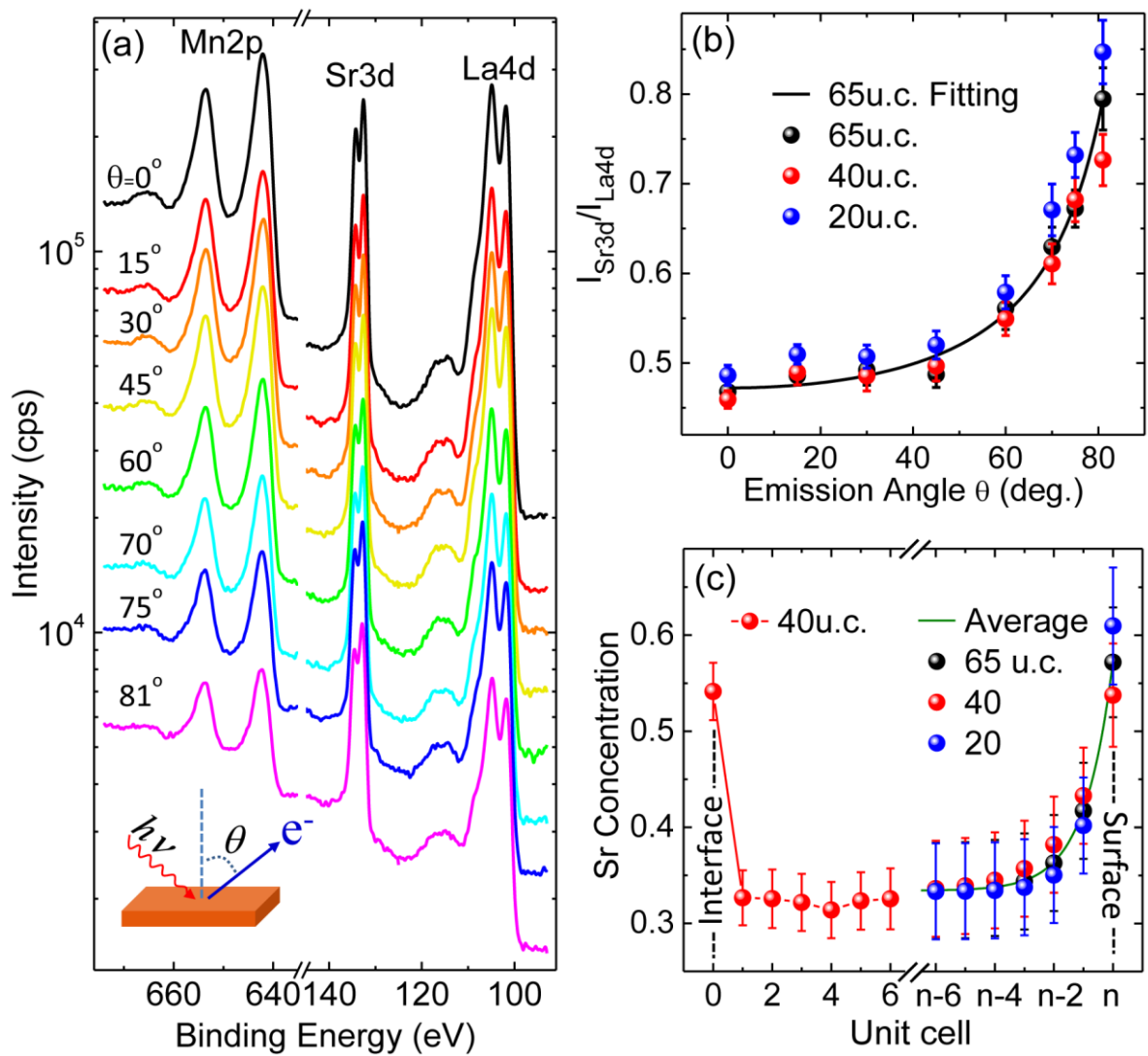

**Figure 3** (*Color online*) (a) ARXPS spectra of a 65 u.c. LSMO film with an inset showing the schematic of ARXPS experimental setup. (b) The experimental (20, 40 and 65 u.c.) and fitted (65 u.c.) intensity ratios of Sr3d/La4d as a function of emission angle for LSMO films. (c) Layer-by-layer dependence of Sr concentration of LSMO thick films at the interface (*left*) determined by STEM/EELS and near the surface (*right*) determined by ARXPS.

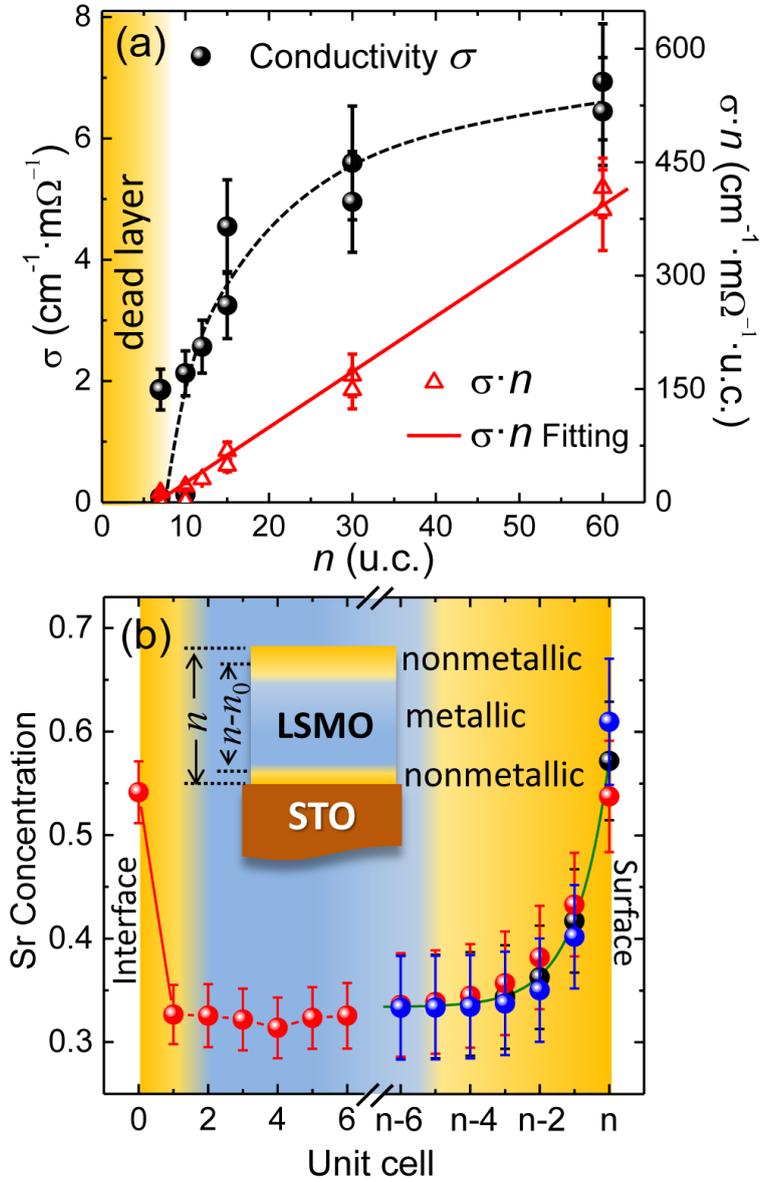

**Figure 4** (*Color online*) (a) Thickness dependence of measured conductivity $\sigma$ and $\sigma \cdot n$ for LSMO films for the thickness $n > n_{cr} = 6$ u.c. at 6 K. The solid data points are obtained from the films with thickness n ≥ 7 u.c., and dashed line is a guide to the eye. The solid red line is the linear fitting to data by assuming $n_0$ as the nonmetallic layers existing at the surface and interface, (b) Schematic drawing of nonmetallic layers (yellow) and the conducting layers (blue) superimposed on the data from Fig. 2 with the inset as a model structure.

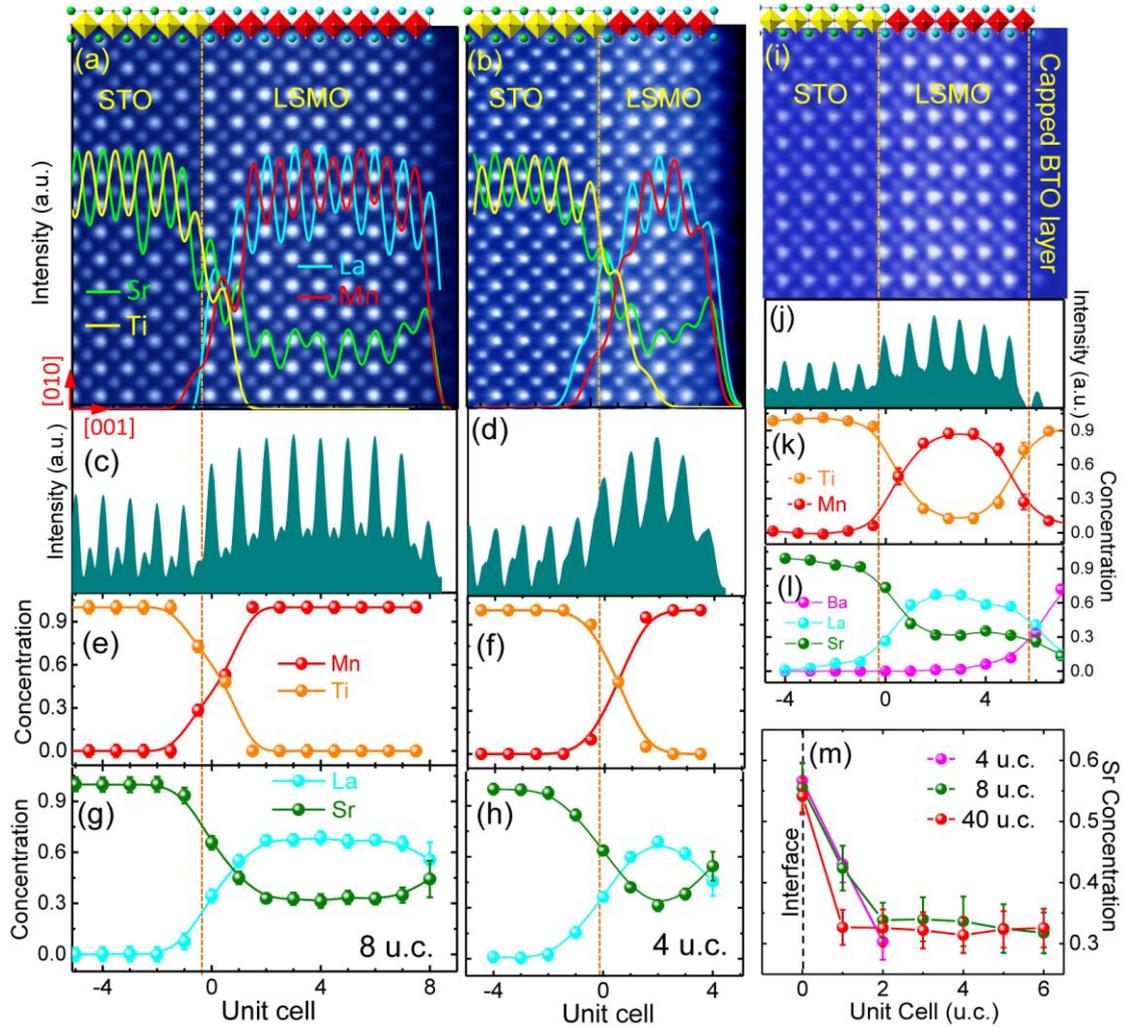

**Figure 5** (*Color online*) HAADDF-STEM images and EELS profiles of (a) 8 u.c and (b) 4.u.c. LSMO films along [100] direction, respectively. (c), (d) The corresponding image intensity profile across the STO/LSMO interface. (e) Mn, (f) Ti, (g) La and (h) Sr concentration profiles determined from EELS data. (i) HAADDF-STEM image of 6 u.c. LSMO film capped with BTO and corresponding (j) image intensity profile, (k) Ti and Mn, and (l) Ba, La, and Sr concentration profiles. (m) EELS determined Sr concentration for 4, 8 and 40 u.c. LSMO films.

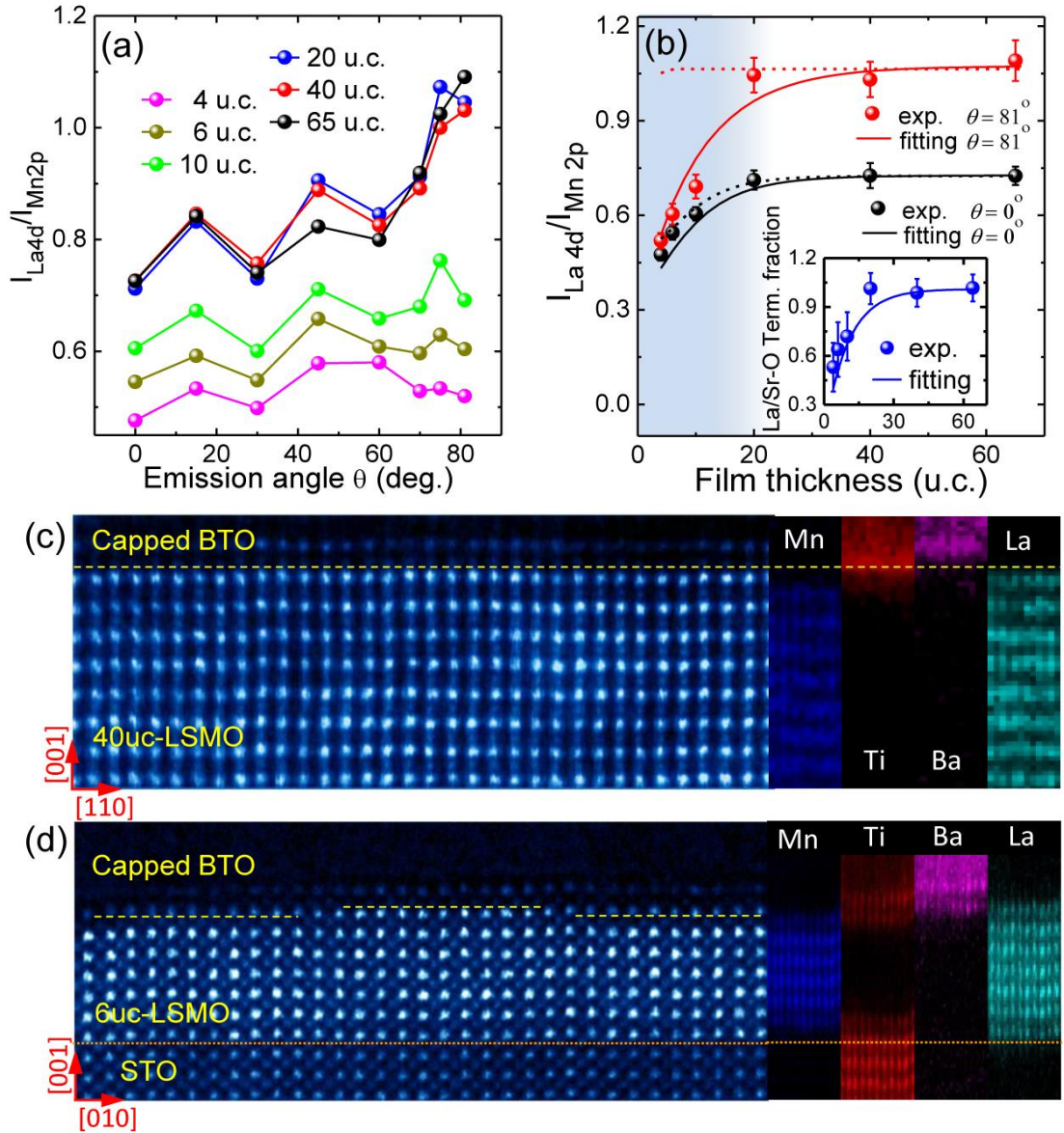

**Figure 6** (*Color online*) (a) Intensity ratio of La4d to Mn2p cores as a function of the emission angle $\theta$ for different thickness of LSMO films. (b) Intensity ratio of La4d to Mn2p core as a function of film thickness for $\theta = 0°$ and $81°$ compared to the fitting results with (solid curve) and without (short dashed curve) mixed termination. The inset presents the determined fraction of surface La/Sr-O termination for different thickness of LSMO films. The shaded area denotes the thickness region of the films with mixed $MnO_2$ and (La/Sr)O termination. HAADF-STEM image in the proximity of the surface of (c) 40 u.c and (d) 6.u.c. LSMO film overlapped with elemental-specific EELS spectroscopic mapping. The films were capped with amorphous $BaTiO_3$ deposited at room temperature to protect the film surface which was marked by yellow dashed lines.